\documentstyle[12pt,aasms4]{article}
\begin{document}
\title{Near Infrared imaging of the proposed z=2 cluster behind 
CL 0939+4713 (Abell 851)}
\author{J.B.Hutchings}
\affil{Dominion Astrophysical Observatory, National Research Council of
Canada, 5071 W. Saanich Rd, Victoria, B.C. V8X 4M6, Canada;
hutchings@dao.nrc.ca}

\author{T. J. Davidge}
\affil{Canadian gemini Project Office, Herzberg Institute of Astrophysics, 
5071 W. Saanich Rd, Victoria, B.C. V8X 4M6, Canada; davidge@dao.nrc.ca}

\centerline{and}

\affil{Department of Physics and Astronomy, University of British Columbia, 
Vancouver, B. C. V6T 1Z4 Canada}

\authoremail{hutchings@dao.nrc.ca,davidge@dao.nrc.ca}

\begin{abstract}

	We have obtained deep J and K' images of a 90 arcsec square field 
containing the z=2.0 QSO that is viewed through the z=0.4 rich cluster A851. 
Brightnesses of individual galaxies are measured from these data, and archival
F702W WFPC2 images from HST. The results are used to 
construct 2-colour diagrams, colour-magnitude diagrams, and luminosity 
functions of galaxies in A851 and the faint objects discovered by Dressler
et al (1993) near the QSO. The QSO faint companions appear to form a separate
population from other faint galaxies in the field. 
Comparisons with the GISSEL models covering redshifts from 0 to 3 
indicate that (1) the A851 galaxies have ages between 3 and 10 Gyr, and 
(2) if they are at the QSO redshift, the QSO companion galaxies are within 1Gyr
of cessation of star-formation. 
\end{abstract}

\section{Introduction and Observations}

  While studying the z=0.4 rich cluster Abell 851 (0939+4713) with HST images, 
Dressler, Oemler, Gunn, and Butcher (1993: DOGB) detected an apparent 
cluster of compact faint galaxies around a QSO with z=2.06. They
suggest that these galaxies may be in the very early stages of formation, and
thus contain important clues for galaxy evolution. If these galaxies are 
truely young then they will have a very blue 
spectral energy distribution (SED); conversely, if a substrate of 
old stars is present then it should be seen at red wavelengths.

  The discovery HST images were taken with the uncorrected optics of WF/PC.
Since then, the field has been re-observed with the corrected WFPC2 camera,
with the area of interest in one of the WF detectors (0.1 arcsec pixels). The 
WFPC2 exposure, taken through the F702W filter, totalled 21000 seconds; we have
obtained the data from the HST archive, and generated a summed image cleaned
of cosmic rays. In the current paper we couple photometric measurements from 
this HST image (which we refer to as `R') with those obtained 
from a deep near-infrared survey of the field around the z=2 QSO. 
Our primary objective is to obtain colour-colour plots of the faint objects 
discovered by DOGB, and investigate their evolutionary state and redshift 
by making comparisons with state-of-the-art stellar population 
models. Since the field lies in the rich z=0.4 Abell
851 cluster, we will also have data for its members. Given that the redshift of
the main cluster is known, the photometry of cluster members provides a means 
of assessing the success of the models.
Stanford, Eisenhart, and Dickinson (1995) published NIR imaging of the
A851 cluster, which is not as deep as ours, and is not useful for the
z=2.05 QSO region galaxies. However, their study provides a good reference 
for the brighter z=0.4 cluster galaxies.

   The deep J and K' images of the z=2.0 QSO and its possible protocluster 
were recorded in January 1996 with the Canada France Hawaii 
Telescope Redeye camera, which was equipped with the wide-field optics. 
The images were sampled with 0.5 arcsec pixels and the point source FWHM was
measured at $\sim$0.8 arcsec. Total integration times were
32 minutes in each filter, taken in a 4 x 4 grid with 10 arcsec
steps; the individual exposure times were 2 mins in J and 30 sec in K'. 
Sky flats were obtained by median-combining the rastered frames. 
The flat-fielded images were registered and then co-added, and the fully 
exposed field size is 90 arcsec. Standard stars were observed for photometric
calibration, and the sky was clear. The instrumental K' observations were 
transformed into K magnitudes. 

\section{Measurements}

    Figure~\ref{dresa} shows the 3 images of the field. Brightnesses of
individual galaxies were measured using two different 
techniques. First, the IRAF task `imexam' was
used on all marked galaxies with galaxy and background windows adjusted
individually to allow for the range of sizes and crowding. Several different
measures were made of each galaxy to confirm convergence on the best value.
Second, brightnesses of the extended galaxies thought to be members of 
A851 were measured within a fixed 3 arcsec radius aperture 
using the PHOT task in DAOPHOT (Stetson 1987). This was done using the
WFPC2 image smoothed to the NIR resolution, and galaxies were selected
only if measurable in all three colours. Effectively this amounts to K-band
selection of the faint end.  
Comparison of the two methods of measurement for objects in common 
showed good agreement (no zero point or ratio differences). 
In all, we have photometry in R, J, K for $\sim$140 galaxies in the 
fields sampled by our NIR imaging. The errors in the photometry depend
on the signal level, sky brightness, and detector read noise. From the values
in our data we calculate 1$\sigma$ 0.1 magnitude errors at $K = 18.8, J = 20.3$, 
and $R = 23$. (The difference between our two sets of measures is well within these, 
as expected since they have the same photon noise.)
Table 1 shows the photometric measures of the galaxies identified by DOGB.
We have included several more faint galaxies within the same area in our
QSO group. These and all other measures are available in machine-readable
form from the authors.

    Before considering the photometric properties of the objects in the images, 
we first consider the statistical significance of the faint objects detected 
near the QSO. To do this, galaxy counts were made in 20 arcsec square boxes 
at various locations in the HST images. These measurements were made from the
HST image, as it is deeper, less noisy, and better able to resolve closely 
spaced galaxies than the NIR data.
The box centred on the QSO contains 2 bright galaxies and 24 faint galaxies.
Boxes at other locations had counts of 2.7 bright, and $12 \pm 2.7$ faint
galaxies, which we define to be those objects with R$>$23. The bright 
galaxies are large and are either members of the z=0.4 main 
cluster or the foreground. These measurements show that the
QSO subfield has a faint galaxy excess at the 3-4$\sigma$ level. 
Thus we confirm that there is an excess of faint galaxies 
near the QSO, that is almost certainly a background cluster. 
DOGB also discuss statistics that indicate the cluster is real. Such clustering
has been seen around other z$>$2 QSOs (see Hutchings 1995 and references
therein), and may be attributed to foreground 
lensing that leads to the discovery of such QSOs, or a real group 
associated with the QSO. In this instance the QSO discovery was made by random
sampling within the Abell 851 field by Dressler and Gunn (1992), so 
selection by lensing is unlikely.

   If the faint objects near the QSO are more distant than other faint 
objects in the field, then they might be expected to have 
different sizes and degrees of compactness. To determine if this is the 
case, key properties of the faint galaxies near the QSO were compared 
with those detected elsewhere in the field. For the purposes of this 
comparison we measured for each galaxy the FWHM of the azimuthally averaged
flux profile, the peak WFPC2 signal, and the total flux within a 5 arcsec 
diameter. The statistics for the QSO faint companions were compared with those 
for similar numbers of faint  (R$>$23) galaxies elsewhere in the field,
with the same distribution of brightness. The latter groups were drawn from the 
highest concentration of such objects in a similar area, as well as from random
places. We then compared the distributions of the above measures in the three
sets of galaxies. We find that the distribution of peak WFPC2 
signal for the QSO companions differs from the two comparison faint galaxy sets
at the $\sim$3$\sigma$ level, being overall higher by a factor 2.6 in the mean. 
Since the fluxes are matched, the ratio of peak signal to total flux (i.e. the
compactness of the objects) is higher -- at the 2-3$\sigma$ level, 
with the distribution for the QSO companions being 35\% more compact by this measure. 
Finally, the galaxy size down to the detection limit has a wider distribution 
among the QSO companions.
Thus, the QSO companions are different from other faint galaxies in the field 
by being centrally brighter, more compact, and showing a wider range of size.

   Figure~\ref{dresb} shows the magnitude distributions of the measured 
galaxies in all three colours, for the general z=0.4 cluster field and for the 
QSO companions as defined by DOGB. We show only galaxies that are measured
in all three colours. The distributions for both sets of galaxies are similar,
and show peaks near the same magnitudes. The histograms rise sharply by a 
factor 2-3 at magnitudes fainter than R=23, J=21, and K=19.5, and the QSO 
group has no member brighter than this (except 
the QSO itself). The mean magnitudes of the QSO group are on average
0.4 mag brighter than the faint galaxies in the rest of the cluster field
(R$>$23, as described above).
Considering the selection effects in choosing the QSO group, these may
not be significant differences. If they are at z=2, they have brighter
absolute magnitudes than the galaxies in the z=0.4 cluster.

\section{Models and data plots}

   We have generated a suite of models with the GISSEL code 
(Bruzual and Charlot 1993), that represent
the observed R, J, and K' magnitudes and colours of galaxies undergoing a
1 Gyr starburst, followed by passive evolution. 
These are shown in Fig~\ref{dresc} for a range of redshifts. 
The addition of dust in the rest frame moves the tracks along
a reddening line that is nearly parallel to the unreddened sequences. 
Of the various possible combinations, the R-J/J-K 2-colour diagram 
provides the best separation of different redshifts. While the
positions of tracks on this diagram at different redshift 
overlap somewhat, the age distribution along the tracks are very different.
The tracks run to ages of 20 Gyr where complete. At higher redshift,
such ages are of course not possible, but equal age intervals move them
across the diagram faster, since we are observing the SED evolution further 
into the rest UV. The tracks show ages at 1 Gyr intervals. 

If star-formation 
is continuous or renewed, then galaxies move more slowly or move back
along the same tracks.  At redshift 2, the largest
reasonable age for the universe is about 3.5 Gyr (H$_0$=50, $\Omega$=0.2).
Continuous star-formation, or exponentially declining star-formation over Gyr
timescales, gives maximum J-K of $\sim$1.4 and R-J of $\sim$1.3 at this age. 
This is marked with an asterisk in Fig~\ref{dresc}. The fastest possible tracks 
(very short initial starburst) reach J-K=2.5 and R-J=7 at age 3 Gyr.
Fig~\ref{dresnew} shows different model tracks for redshift 2. In general,
the lower track is followed as long as star-formation continues. When star-formation
stops, the tracks evolve more rapidly to redder R-J values. Thus the diagram
shows two of the family of such tracks, and others can be interpolated by eye.

   We consider 3 subsets of the photometric measurements. First, we select 
the faint galaxies near the QSO. Second, in order to eliminate objects with 
large measuring scatter, we select all other galaxies in the field with measuring
errors less than 0.3 mag in all three colours. This will largely comprise the
z=0.4 cluster galaxies. 
Third, we select all faint galaxies (R$>$23) in the field not near the QSO.
These have the same R, J, K magnitude range as the QSO companions. This is used
as a faint galaxy comparison control sample.

   We show the distribution of the galaxies superposed on the tracks in 
Fig~\ref{dresc}.
The main cluster population forms a very well-defined group, and we have
adjusted the model tracks arbitrarily in J-K (by less than 10\%) to match
this sequence. The bandpasses and systematics in our photometry are uncertain by
this amount. R-J is less uncertain and is not adjusted. We also note that
our measured colours are in good agreement with the cluster galaxies
measured by Stanford et al (1995). In Figure~\ref{dresc} 
the main population of z=0.4 cluster galaxies is seen at a position suggesting
ages in the range 3-10 Gyr, with no reddening.

  The field may contain galaxies at any redshift, and it would be nice to
use a two-colour diagram to separate them. However, as noted above, the tracks for all
redshifts from 0.4 to 2 lie close together, and separate out only in
the distribution of galaxies along the different tracks with age. At 
redshift 2, post-starburst galaxies evolve rapidly to higher value of R-J,
but continued star-formation keeps the colours bluer than those observed
in the QSO faint companions, as noted in Fig~\ref{dresnew}. 
In this diagram, we show the QSO companion galaxies, and it is seen that as a group
they correspond to populations shortly after the end of an initial starburst
of duration 1-2 Gyr, if they are at the QSO redshift of 2. In Figure~\ref{dresc}
we see that the position of these galaxies corresponds to populations which 
become younger as
their redshift increases. Thus, for redshifts 1.0 or higher, the
galaxies have populations defined by stars less than 2 Gyr after the
end of star-formation.

   The control faint galaxy plot (Figure~\ref{dresc}) 
has a similar distribution, with wider
J-K colour range. These measures were done without special attention to
contamination by nearby companions in the NIR images, so more scatter may be
expected. We note that the faint galaxies in our control sample 
typically have R=24, which corresponds to $M_R \sim$-17 at
z=0.4 for H$_0$=80; consequently, some of these objects may be dwarf galaxies
in the z=0.4 cluster. Some of the control galaxies have redder J-K colours,
and may be high redshift (z$>$2) objects in active star-formation. Thus,
there are some that appear to belong to the main cluster
population, and possibly some very high redshift galaxies. The QSO
companions have slightly higher extreme R-J values, suggesting as noted above, that
their star-formation has ceased. Overall, the distributions in the two panels of
Figure~\ref{dresc} suggest they are not the same galaxy populations, and that
the QSO companions are a distinct group. Thus, our 2-colour
photometry suggests (but cannot prove) that the QSO companions are either 
faint members of the z=0.4
cluster, or are immediate post-(1-2 Gyr)-starburst objects at z=2.

   In Figure~\ref{drese} , 
we show the galaxies plotted on a colour-magnitude diagram.  This diagram shows
the same bright galaxies, QSO companions, and the faint control group all together. 
The diagram illustrates the bright cluster sequence and also that some of the faint
galaxies belong to other populations. In order to relate the colours to
redshifts and ages,
the diagram also includes models showing how the same (R-J) colour index changes
with population age (indicated along the top), and redshift, as labelled.
The diagram does not imply that the ages and magnitudes are related. It does
allow the reader to look at measured objects and see what combination of
redshift and population age they may correspond to. Note particularly how R-J 
increases when the star-formation ends. At z=2, continued star-formation keeps
R=J below 1.3. The bright galaxies are consistent with unreddened z=0.4 objects
with populations older than 3 Gyr. The QSO 
faint companions have a colour spread that is similar to the control galaxies
but reach slightly higher R-J values. At z$>$0.4, most of the `QSO companions'
must have stopped star-formation. The CMD thus further illustrates 
the feasibility of the hypothesis that the QSO companions are associated with 
the QSO if they are immediate post-starburst objects.

   This raises the question of whether the QSO phenomenon might cause the local
star-formation to halt (by radiation effects?), or whether the QSO phenomenon
arises subsequent of the cessation of star-formation in the group (stripping
of gas by merging?). Finally, we may examine whether the QSO faint companions may
evolve into a known present-day counterpart. At magnitude 23 and z=2, and with their
supposed stellar population, they correspond to absolute magnitude R$\sim$-21.
The zero redshift evolution of these populations (seen in Fig~\ref{dresc})
has little colour change, so that they may evolve passively into present-day
bright galaxies.

\section{Summary}

   We have shown that the cluster A 851 is populated by galaxies of ages
3 to 10 Gyr, with no reddening. This applies to the cluster population ranging
over 4-5 magnitudes and is more than 2 magnitudes deeper than the measures
reported by Stanford et al (1995). Fainter than this there is an excess of
galaxies which may be  
at higher redshifts, dominated by younger populations. If they are at redshift 2, the
QSO companion galaxies appear to be within 1 Gyr after an initial (1-2 Gyr) starburst. 
The spread of other faint galaxies in the field suggests that they have
a different distribution of redshifts and ages.
However, it is impossible to determine definitively whether there is a
cluster associated with the z=2 QSO.

   Spectroscopy is needed to study these particular galaxies further. Observations
of the kind described in this paper would be better able to distinguish galaxies 
at redshifts larger than 2, as the J-K changes rapidly in this redshift range, 
for all population ages. 
We note that the z=2.37 QSO 0820+296 appears to be such a case (Hutchings and
Neff 1997).
  
    We thank Susan Neff for participating in obtaining the CFHT observations.
We thank the referees for their careful attention to these conclusions.
  
\clearpage
     
\centerline{\bf References}

Bruzual G., Charlot S., 1993, ApJ, 405, 538

Dressler A., Gunn J.E., 1992, ApJS, 78, 1

Dressler A., Oemler A., Gunn J.E., Butcher H., 1993, ApJ, 404, L45 (DOGB)

Hutchings J.B. 1995, AJ, 109, 928

Hutchings J.B., and Neff S.G. 1997, AJ (in press)

Stanford S.A., Eisenhart P.R.M., Dickinson M., 1995, ApJ, 450, 512

Stetson P.B., 1987, PASP, 99, 191

\clearpage

\centerline{\bf Captions to figures}

\figcaption[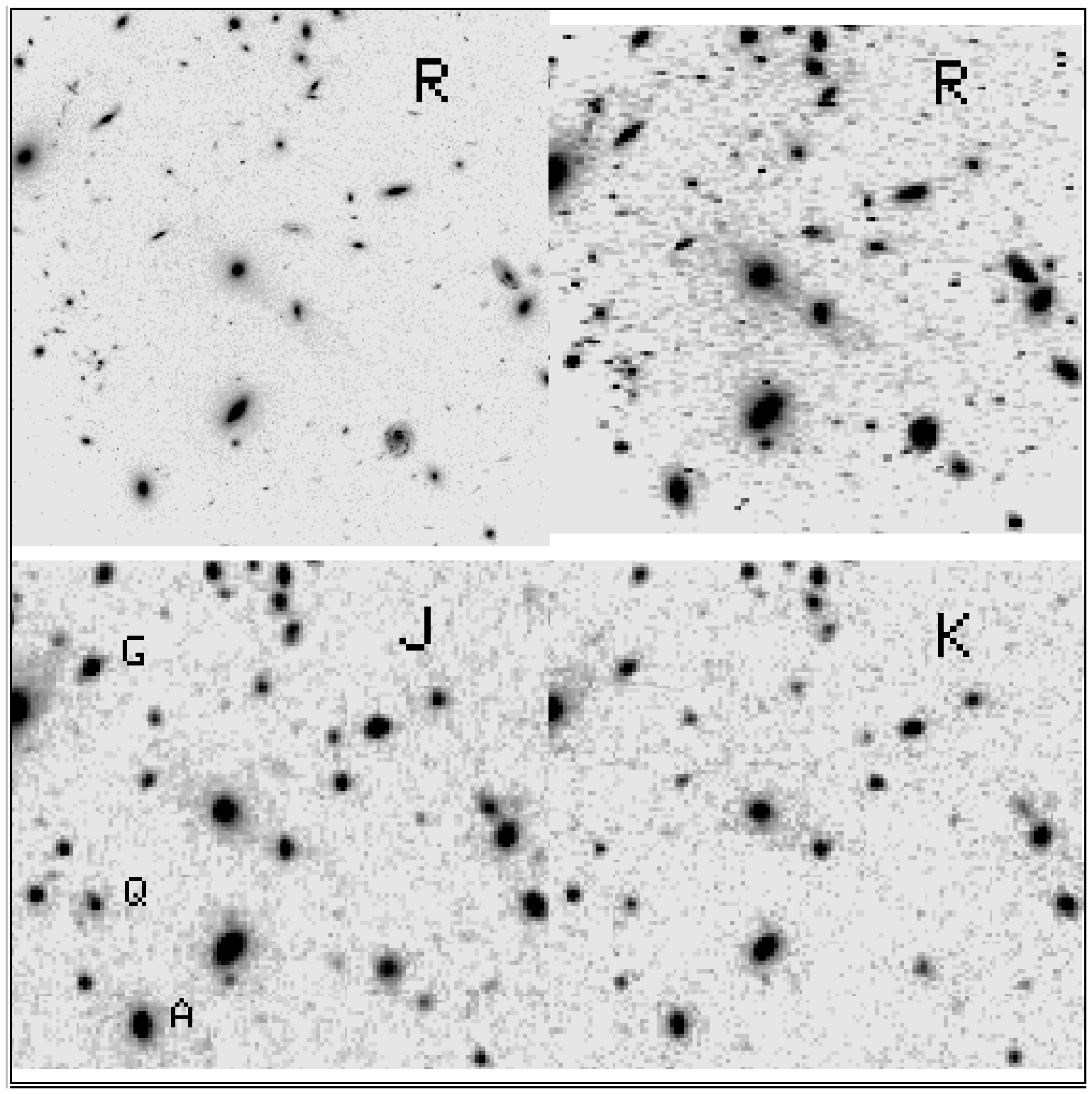]{Images of Abell 851 field.
The top panels show the HST images with original resolution and
smoothed to match the NIR images. The lower panels are the new
J and K images. \label{dresa}}

\figcaption[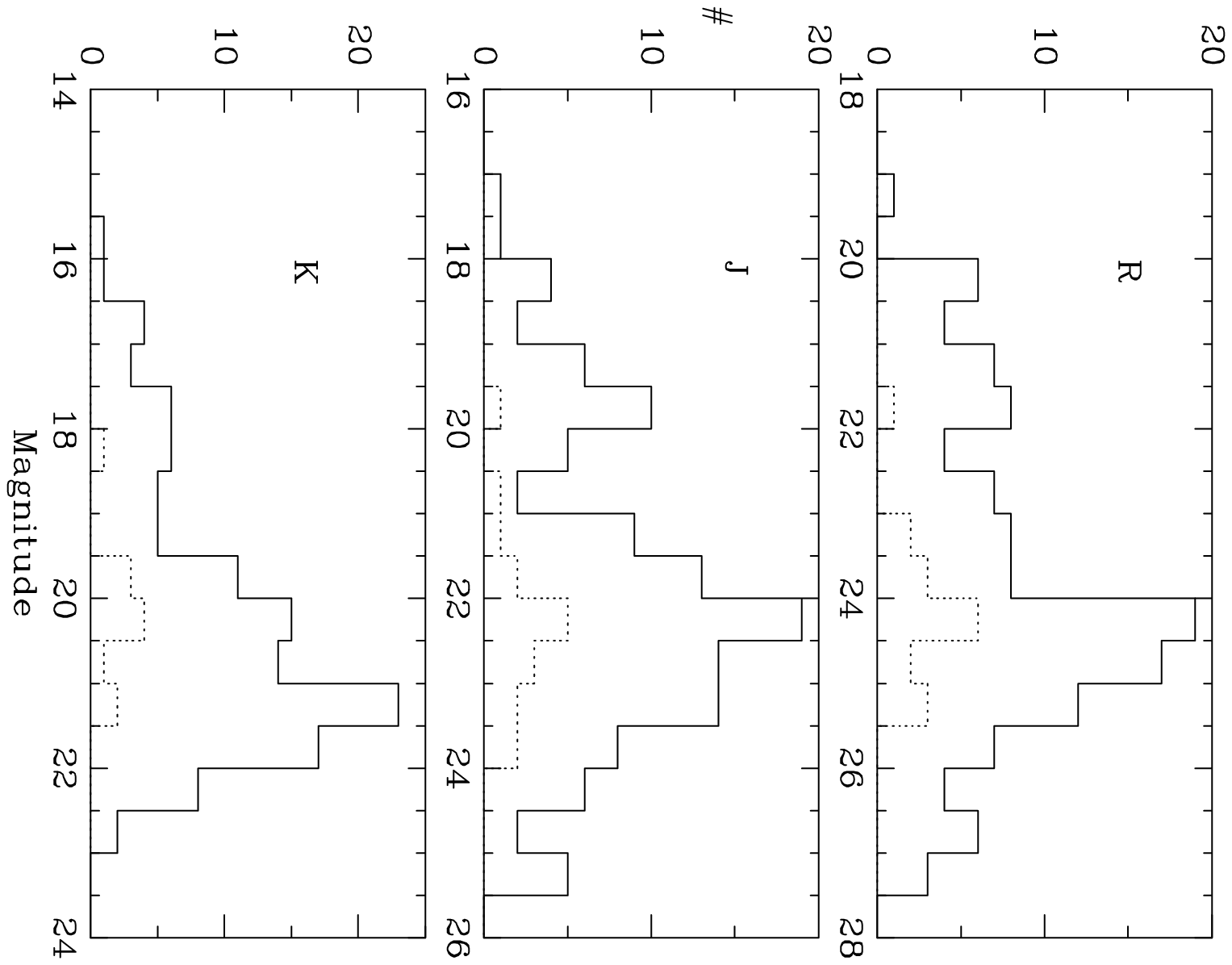]{Magnitude distributions of all measured galaxies 
in 3 colours. Upper envelope is entire field and lower (dotted)
is the QSO and group only. \label{dresb}}

\figcaption[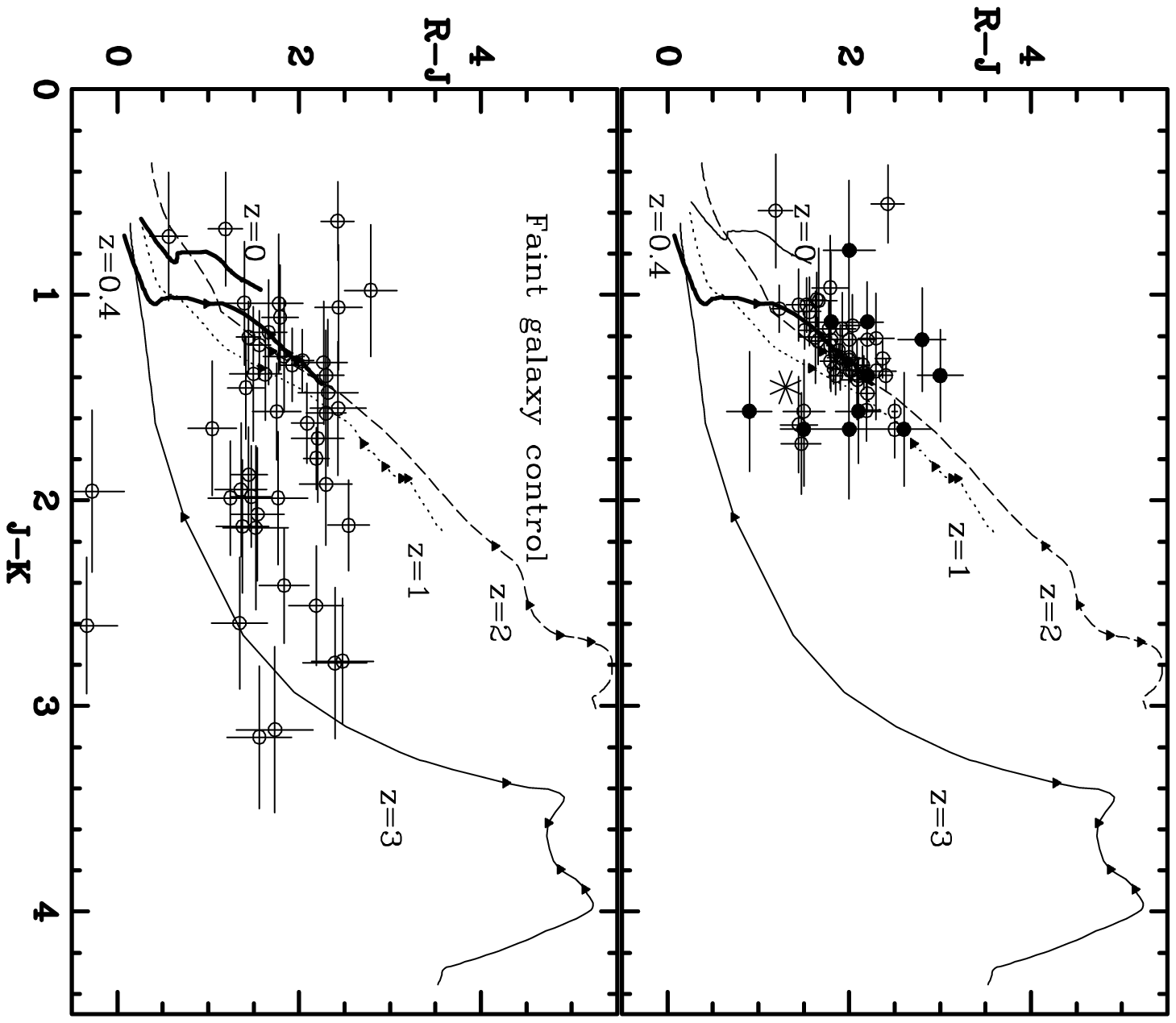]{2-colour diagrams with models.
Lines show evolution of 1 Gyr starburst populations
followed by passive evolution, for redshifts as shown. Small triangles
on the tracks mark positions at ages 1-5 Gyr, and tracks start at lower
left at age 10$^7$ yr. Reddening moves tracks almost parallel to their 
direction to upper right. Upper panel shows bright galaxies from the
whole field (trimmed at K=21, J=22, R=24)
as open symbols, and the QSO and `companions' as filled symbols. 
The QSO is the central point of this group, situated on the thick line.
Lower panel shows  control sample of faint 
galaxies in the field other than the QSO group, but with similar
magnitudes. 
 \label{dresc}}

\figcaption[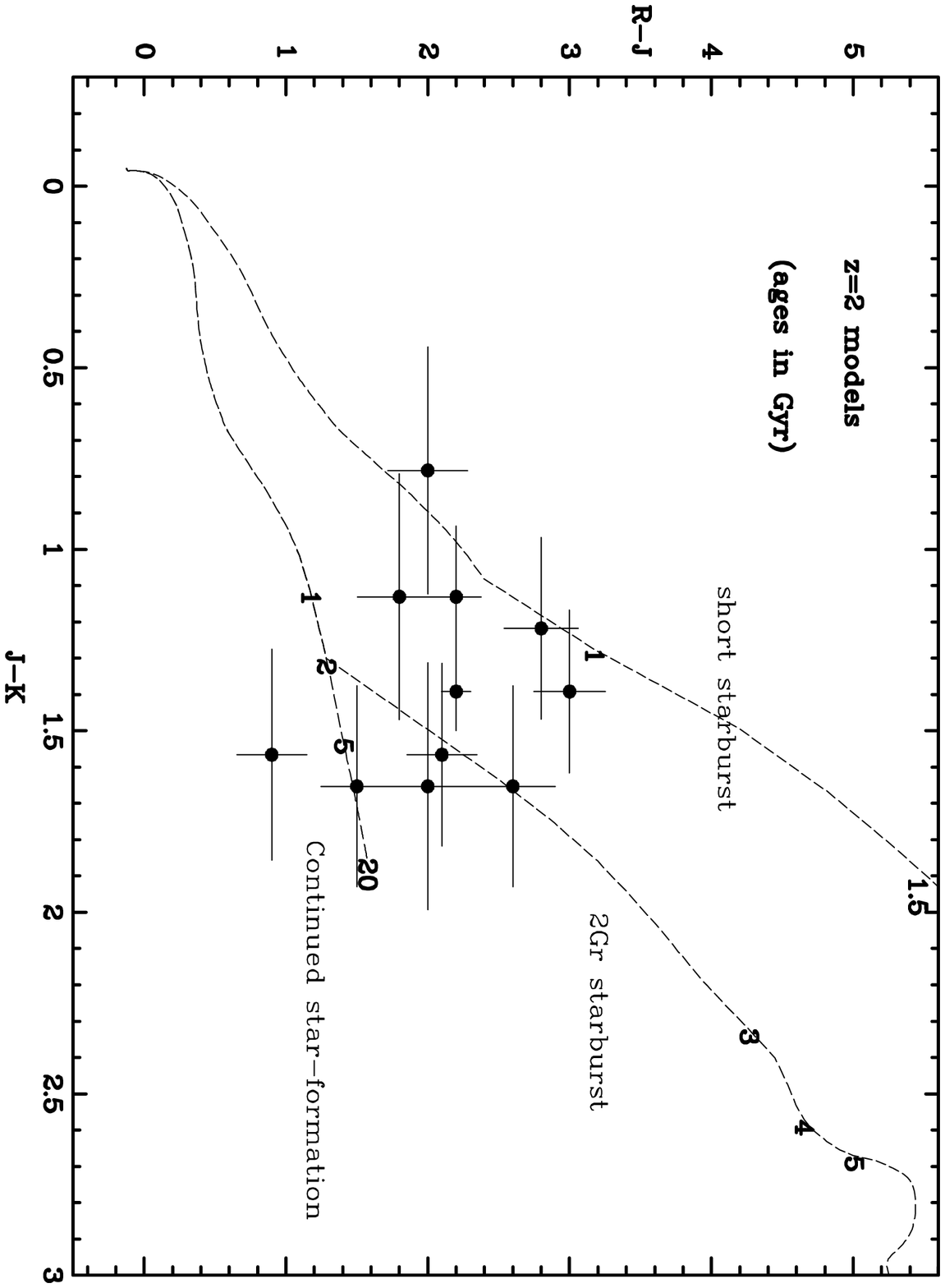]{Models for z=2 with different durations of star-formation,
together with the QSO faint companion galaxies from Fig~\ref{dresc}. Note that
if these galaxies are at z=2, they are post-starburst objects, but with ages
1-2.5 Gyr.\label {dresnew}}

\figcaption[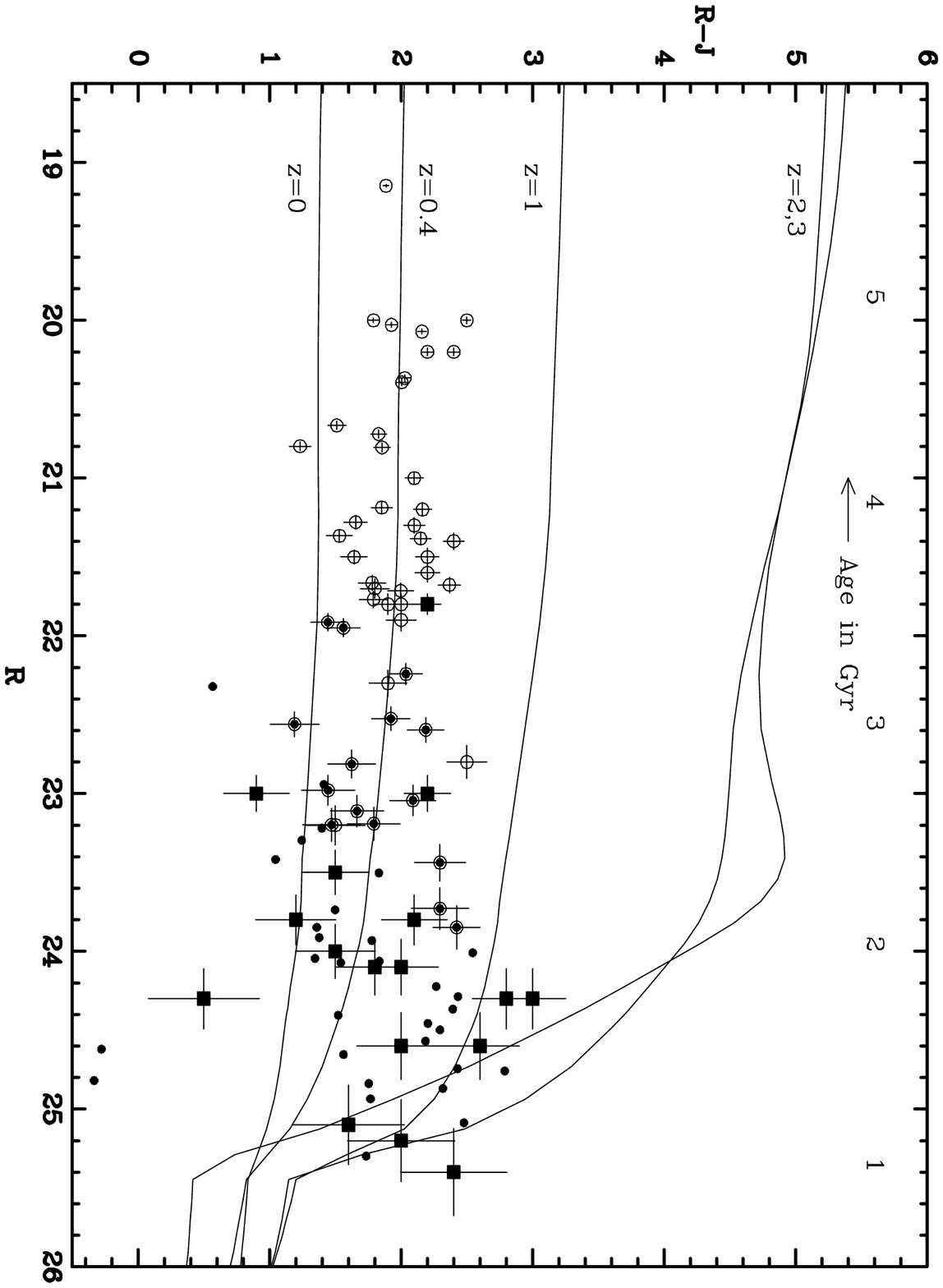]{Colour-magnitude diagram showing bright galaxies as open
symbols, QSO group as closed symbols, and control faint galaxies as small dots.
Error bars are omitted for the latter but all vary systematically across 
the diagram. The lines show this colour evolution with age as shown, along the
top axis, for models at redshifts
shown. Age is not simply related to magnitude, so superposition is only
suggetive. R-J is an age measure if the redshift is known. \label{drese}}

\end{document}